\RequirePackage{amsthm}
\documentclass[pdflatex,sn-mathphys-num]{sn-jnl}% Math and Physical Sciences Numbered Reference Style 
\usepackage{anyfontsize}%
\usepackage{lineno}%
\usepackage{graphicx}%
\usepackage{multirow}%
\usepackage{amsmath,amssymb,amsfonts}%
\usepackage{amsthm}%
\usepackage{mathrsfs}%
\usepackage[title]{appendix}%
\usepackage{xcolor}%
\usepackage{textcomp}%
\usepackage{manyfoot}%
\usepackage{booktabs}%
\usepackage{algorithm}%
\usepackage{algorithmicx}%
\usepackage{algpseudocode}%
\usepackage{listings}%
\usepackage{soul}%
\begin{document}

\title[Article Title]{Neuromorphic Infrared Fibre-Optic Event-Based Sensing with Fast and Efficient Photonic-Electronic Spiking Neurons}

%%\title[Article Title]{Neuromorphic Event-Based Optical Fibre Sensing With A Spiking Resonant Tunnelling Diode}

\author[1]{\fnm{Dylan} \sur{Black}}

\author[1]{\fnm{Giovanni} \sur{Donati}}

\author[1]{\fnm{Joshua} \sur{Robertson}}

\author[2]{\fnm{Qusay} \sur{Raghib Ali Al-Taai}}

\author[3]{\fnm{Jos\'{e}} \sur{Figueiredo}}

\author[2]{\fnm{Edward} \sur{Wasige}}

\author[4]{\fnm{Bruno} \sur{Romeira}}

\author[1]{\fnm{Antonio} \sur{Hurtado}}

\affil*[1]{\orgdiv{Institute of Photonics}, \orgname{University of Strathclyde}, \orgaddress{\street{99 George Street}, \city{Glasgow}, \postcode{G1 1RD}, \country{United Kingdom}}}

\affil*[2]{\orgdiv{High Frequency Electronics Group}, \orgname{University of Glasgow}, \orgaddress{\street{Kelvin Building}, \city{Glasgow}, \postcode{G12 8QQ}, \country{United Kingdom}}}

\affil*[3]{\orgdiv{LIP and Dep. Física, Faculdade de Ciências}, \orgname{Universidade de Lisboa (FCUL)}, \orgaddress{\postcode{1749-016} \city{Lisboa}, \country{Portugal}}}

\affil*[4]{\orgname{International Iberian Nanotechnology Laboratory}, \orgaddress{\street{Av. Mte. José Veiga s/n}, \city{Braga}, \postcode{4715-330}, \country{Portugal}}}

\abstract{Current photonic technologies for remote sensing require the capture of large amounts of data, suffering as a result from high energy consumption, excessive data redundancy and storage, and costly data processing requirements, limiting their ability for direct and efficient edge-processing for rapid decision making and alarm triggering. In contrast, biological sensing systems, given their event-driven nature and in-sensor processing capabilities offer energy-efficient and practical alternatives. In this work, we draw direct inspiration from the neural spiking in biological sensory systems, to propose a novel neuromorphic event-based photonic technology permitting the remote sensing of environmental events-of-interest efficiently, at high speeds and across a wide dynamic frequency range. Our approach combines widely-deployed fibre-optic telecommunication links and photo-detecting resonant tunnelling diodes (pRTD) acting as light-triggered spiking neurons. We demonstrate that this new neuromorphic photonic sensing approach allows the remote detection of different types of environmental events, including temperature variations, strain-induced motion, audio and air turbulence, with high temporal resolution (encoding them with fast nanosecond-rate neural-like spikes). These results pave the way for novel light-enabled neuromorphic remote infrared fibre-optic sensing networks that are fast, efficient, event-driven, scalable, permit direct processing at-the-edge, and offer practical alternatives to current data-intensive approaches for photonic remote sensing.}


\maketitle

The Information Age, the era we currently live in is characterised by the vast accumulation, processing and storage of data. This data originates from countless embedded sources within our environment and technologies for applications in various fields including autonomous vehicles \cite{Vinoth2024}, perimeter security \cite{Tomasov2025}, weather monitoring \cite{Allen2025} and medicine \cite{Mahato2024}. Among these sources are hardware sensors, devices designed to extract raw data about physical phenomena from their surroundings such as temperature, light, sound and motion \cite{Nwakanma2024}\cite{Fu2020}\cite{Awais2023}. Numerous hardware sensors are employed in modern electronics, including among many others, thermistors for temperature measurements \cite{Labrado2019}, microphones for sound detection \cite{Kumar2022} and photodiodes for sensing light \cite{Kim2021}. Further, leveraging the advantages of photonics for faster response times and expanded operational bandwidth, various optical sensors have been developed for use in environmental and industrial settings \cite{Butt2024}. Among these are sensors that can be implanted into fibre-optic cables, sensitive to temperature and physical strain fluctuations, primarily used for remote environmental and infrastructural health monitoring systems. A widely used type of fibre optic sensor with ample practical field applications is the Fibre Bragg Grating (FBG) sensor \cite{Hill1978}. FBG sensors utilise a distributed Bragg reflector directly inscribed into the optical fibre core, where a periodic change in the refractive index of the material allows specific wavelengths of light to be transmitted, while others are reflected. A key feature of FBG sensors is their sensitivity to environmental changes such as temperature fluctuations \cite{DeTommasi2023} and mechanical strain \cite{Jaime2017}, both of which induce a shift in the peak reflected (Bragg) wavelength of the grating. This high sensitivity to typical physical phenomena, added to their seamless coupling to widely deployed fibre-optic telecommunication infrastructure, highlights FBGs as a key technology for a wide range of remote, light-enabled fibre-optic sensing applications. These include use in structural health monitoring systems used for bridges or buildings \cite{Kinet2014} and biomedicine for use in wearable health diagnostic technologies \cite{Krizan2025}, among others.

\begin{figure}[h!]
    \centering
    \includegraphics[width=\textwidth]{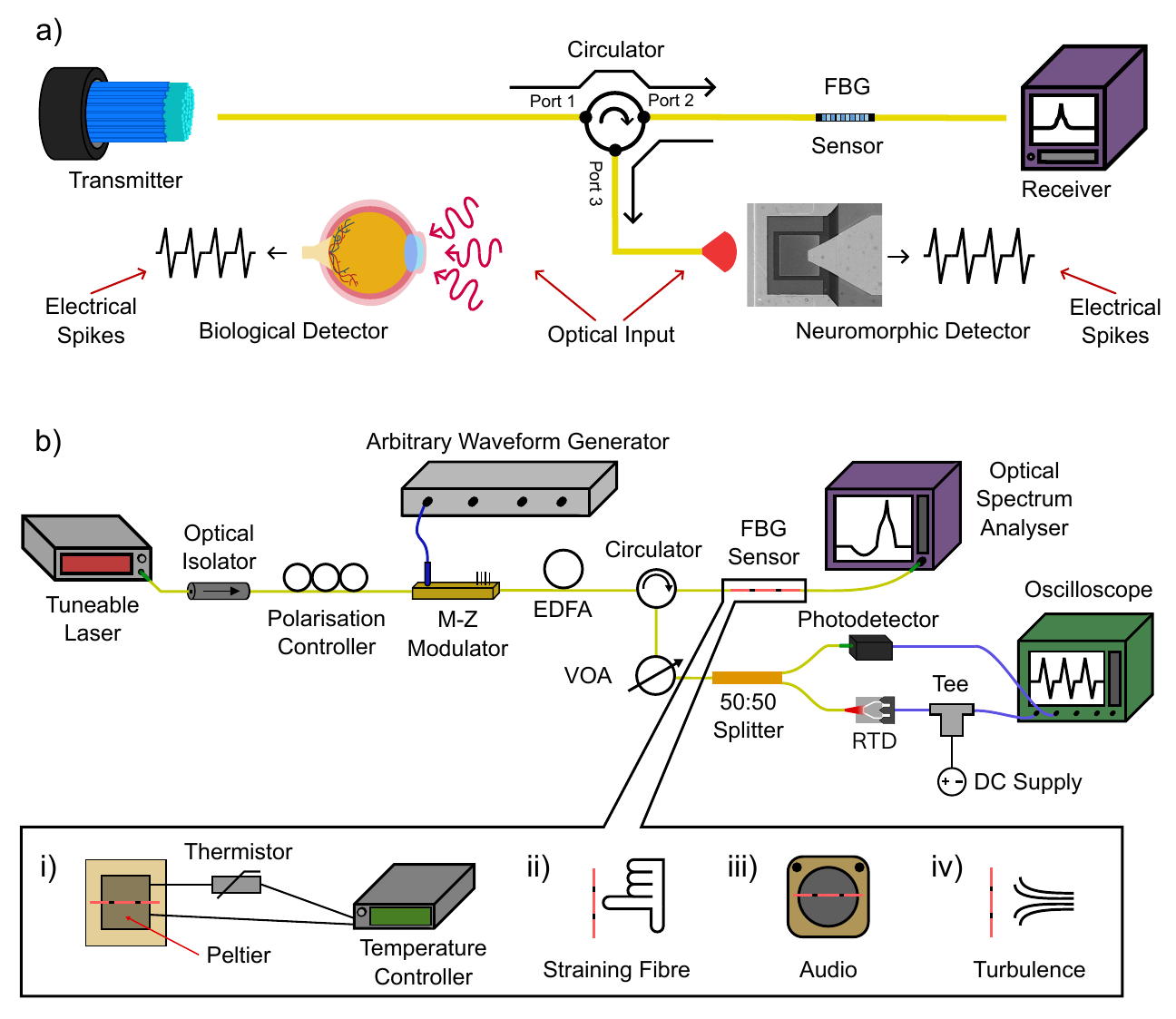}
    \caption{(a) Schematic diagram of the proposed fibre-optic infrared event-based remote sensing technology. A transmitter sends an optical signal to a receiver through a fibre-optic cable, where if any disturbances occur, an embedded Fibre-Bragg Grating (FBG) sensor element will reflect a portion of the light signal to an ultra-fast spiking event-based photo-detecting neuron acting as a neuromorphic photonic interrogator. (b) Experimental setup. EDFA - Erbium Doped Fibre Amplifier, FBG - Fibre Bragg Grating, VOA - Variable Optical Attenuator, RTD - Resonant Tunnelling Diode, DC - Direct Current. (i) - (iv) Equipment change for temperature, dynamic, audio and turbulent measurements.}
    \label{concept_expsetup}
\end{figure}

These photonic sensors perform a similar function to our biological senses, acting as the eyes and ears of existing systems and infrastructure, capturing real-time data about remote environmental conditions which are then sent for analysis and storage by a litany of other hardware processes\cite{Srilakshmi2024}. However, the photonic sensors currently deployed can be inefficient, often capturing large amounts of redundant data whilst needlessly consuming energy and storage. This has led to the rise of nature-inspired neuromorphic (brain-like) sensing hardware approaches, drawing inspiration directly from the energy-efficient and event-based nature of our senses and brain \cite{Wang2025}. In recent years, neuromorphic sensors have seen significant advancements, with the development of the event-based camera (often referred to as a neuromorphic camera) standing out as a major milestone in the field \cite{Lichtsteiner2008}. These cameras use photodetectors to sense changes in (visible) light intensity, encoding the information as asynchronous events that are spike-like in nature, to emulate some features of biological vision \cite{Prophesee2025}. Biologically inspired (neuromorphic) hardware typically uses artificial neurons to generate spiking signals which serve not only as an indication that an event has occurred (switching from an inactive to active state), but additional information about the strength and duration of the signal in the spike rate or temporal characteristics of the spike train. Advancements in neuromorphic technology are firmly rooted in electronics, yet photonics offers solutions to classical electronic bottlenecks using wavelength division multiplexing for highly parallel data transmission to increase operational bandwidth and reducing latency by exploiting the speed of light \cite{PerezLopez2025}. Recently, a convergence has begun to take shape, creating a hybrid technology that leverages the advantages of photonics alongside the infrastructural maturity of electronics, thereby enabling opto-electronic platform development \cite{Tan2022}. Various photonic and opto-electronic candidates have been shown to act like artificial neurons able to generate the same spiking signals used by biological neurons to sense and process information, but at much faster speeds. These have created many lines of intense research for optical spiking platforms based upon different technologies, including semiconductor lasers \cite{OwenNewns2025} \cite{Puts2025}, microring resonators \cite{Biasi2024} \cite{Donati2025b}, memristive devices \cite{Pattnaik2024} \cite{Peng2024}, and opto-electronic oscillators \cite{AlTaai2023} (see \cite{Brunner2025} for a review).

In this work, we propose and demonstrate experimentally for the first time a neuromorphic photonic fibre-optic remote sensing technology, able to operate within the infrared optical communications wavelength range (C-band, centred around 1550\,nm) and using widely deployed fibre-optic telecommunication technologies. This is built with a photo-detecting Resonant Tunnelling Diode (pRTD) \cite{AlTaai2023} to create a novel fibre-based, neuromorphic event-based system capable of producing fast spiking outputs in response to fast environmental fluctuations captured remotely by FBG sensors embedded in fibre-optic links. This process is introduced in Figure \ref{concept_expsetup}(a). Here, a transmitted optical signal is sent to a receiver with an embedded FBG sensor in the optical communication network. If any environmental phenomena occur, such as induced strain or temperature variations to the FBG sensor, the light travelling along the fibre-optic communication link will be reflected towards the neuromorphic photonic-electronic pRTD detector, producing fast spiking signals in response to incoming light with ultrafast response time. The paper is structured as follows: in Section \ref{Results}, we present the results of the neuromorphic pRTD-FBG system in monitoring both slow-varying and fast-changing dynamical disturbances including temperature and strain-induced events involving slow mechanical perturbations at rates of a few \,Hz, audio signals up to 10's of \,Hz and turbulence reaching a few \,kHz. Section \ref{Conclusion} concludes the key findings of the experimental demonstrations alongside some insights on applications of the novel fibre-optic event-based system indicating lines of future research. Finally, Section \ref{Methods} outlines the experimental equipment and layout used for the results presented prior.

\section{Experimental Results}\label{Results}

In this section, we present our experimental results, demonstrating the ability of the neuromorphic event-based sensing system to detect environmental disturbances at differing amplitudes using infrared optical communications platforms. These include temperature variations and strain-induced events on the order of a few Hz as well as faster measurements, including real-time detection of remote RF signals (audio) and fast turbulent events up to kHz rates. The experiments were carried out using the setup shown in Figure \ref{concept_expsetup}(b), with full details outlined in Section \ref{Methods}. 

\begin{figure}[h]
    \centering
    \includegraphics[width=\textwidth]{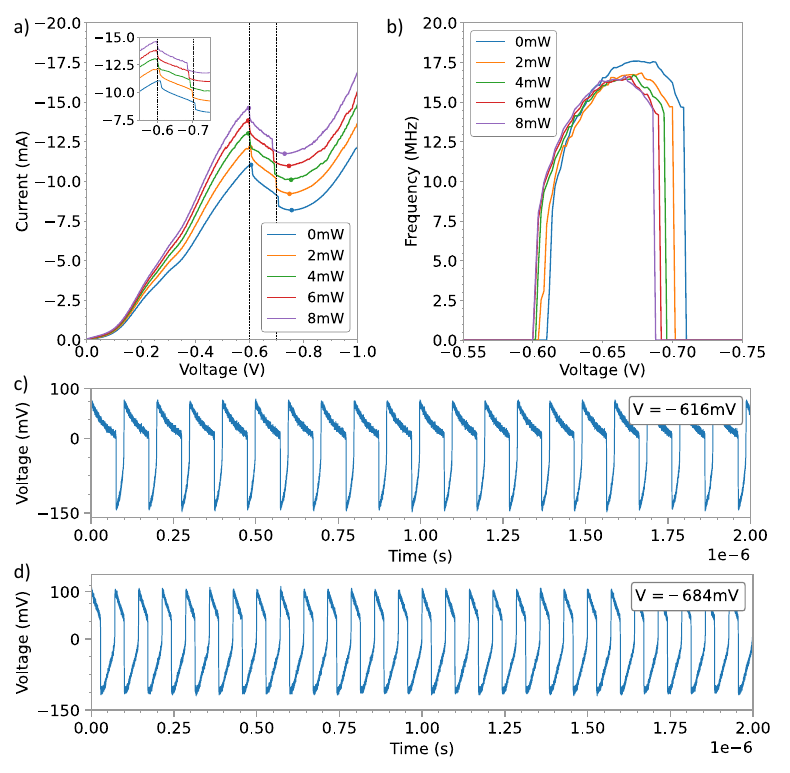}
    \caption{(a) Photodetector resonant tunnelling diode (pRTD) current-voltage (I-V) traces under dark and illuminated conditions with varying optical input powers at the wavelength of 1550\,nm. The insets highlight the shift in the NDR region with increasing light intensity. (b) pRTD frequency-voltage traces for varying intensities of injected optical power, highlighting the variation in spike-firing rate from the system when the optical power is varied from 0 to 8\,mW. (c)-(d) pRTD spiking response when biassed within NDR region with constant voltages at -616\,mV and -684\,mV respectively.}
    \label{rtd characterisation}
\end{figure}

The pRTD has a highly non-linear I-V relationship where for certain applied voltages, the current decreases as opposed to increases as shown in Figure \ref{rtd characterisation}(a). This is due to the presence of a double barrier quantum well (DBQW) embedded within the semiconductor heterostructure, permitting the tunnelling of electrons due to the resonant quantum tunnelling effect, giving rise to a negative differential resistance (NDR) region. Specifically, the DBQW is created from two barrier layers of AlAs 1.7\,nm thick and a well of InGaAs with 5.7\,nm thickness (see \cite{AlTaai2023} for full details). Uniquely, the pRTD devices used in this work contain a layer of photoabsorbing material (250\,nm thick $\text{In}_{0.53}\text{Ga}_{0.47}\text{As}$) permitting the injection of light to generate an additional photocurrent, shifting the I-V characteristics in current and marginally in voltage, as shown in Figure \ref{rtd characterisation}(a), using different colours respectively. For constant applied DC bias voltages close to the onset of the NDR region of the I-V trace, this light induced shift in voltage permits the switching between dark and illuminated states, activating or deactivating a constant electrical spiking regime. The injection of increasing powers of CW light also changes the spike-firing rate across the region of the NDR region as shown in Figure \ref{rtd characterisation}(b). The output electrical spikes are shown for two levels of constant DC bias voltage, highlighted by the dashed vertical lines in \ref{rtd characterisation}(a) at -616\,mV and -685\,mV (as the device operates in reverse bias conditions) as shown in Figure \ref{rtd characterisation}(c)\&(d). These highlight the dynamic change in spiking rate over the range of NDR region, where small stimuli driving the system into the NDR will cause a response with low frequency, while large stimuli driving further into the NDR will cause a larger spiking frequency response.

\subsection{Remote Neuromorphic Photonic Temperature Sensing}\label{Temperature}

\begin{figure}[h]
    \centering
    \includegraphics[width=\textwidth]{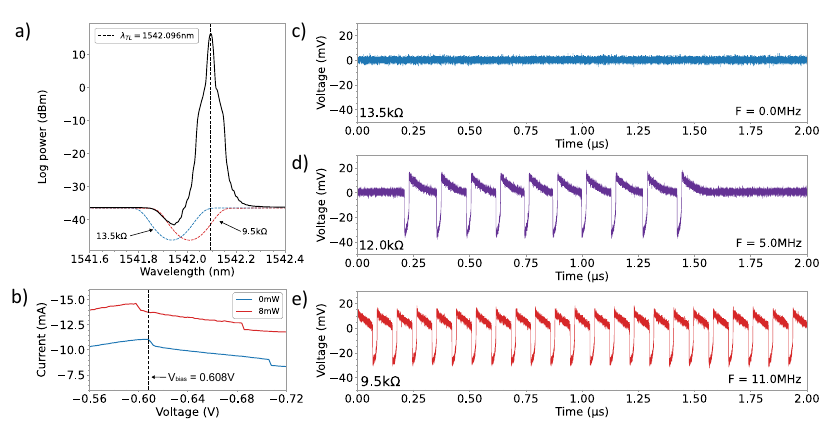}
    \caption{(a) Optical spectra of the tuneable laser, set to emit a wavelength of 1542.096\,nm with the fibre Bragg grating Bragg wavelength shown for 13.5\,k$\Omega$ (blue line, 0\,mW reflected power) and 9.5\,k$\Omega$ (red line, 8\,mW reflected power). (b) Zoom of the resonant tunnelling diode current-voltage trace for a constant biassing voltage of -0.608\,V, highlighting the switch from dark (blue line) to 8\,mW illumination (red line). (c) Oscilloscope readout for a temperature setting of 13.5\,k$\Omega$ (18.22\,\textdegree C), showing no spiking activity (f = 0\,Hz). (d) Oscilloscope readout with an increased temperature setting of 12.0\,k$\Omega$ (20.84\,\textdegree C), showing bursting activity (f = 5.0\,MHz). Final readout for a temperature setting of 9.5\,k$\Omega$ (26.16\,\textdegree C), reflecting the maximum amount of light activating continuous spiking responses (f = 11.0\,MHz).}
    \label{temperatureres}
\end{figure}

At first, we tested the neuromorphic event-based detector to monitor quasi-static environmental temperature variations, using the setup in Figure \ref{concept_expsetup}(b), configuration (i). The outcome of this measurement is presented in Figure \ref{temperatureres}. Firstly, the photo-detecting RTD is (reverse) biased at -0.608\,mV, just before the peak point if its dark I-V characteristic (see Figure \ref{rtd characterisation}(a), blue curve), thus ensuring a quiescent output state under no illumination conditions. In parallel, the tuneable laser was set at a wavelength of 1542.096\,nm and 8\,mW optical emission power, while the thermistor was set to 13.5\,k$\Omega$ (corresponding to a temperature of 18.22\,\textdegree C). This combination sets the system into a regime where the laser emission wavelength is weakly coupled with the FBG sensors Bragg wavelength, as shown in Figure \ref{temperatureres}(a) in the blue dashed line. Hence the reflected light to the photo-detecting RTD is insufficient to make it enter the NDR region of the illuminated I-V curve (Figure \ref{temperatureres}(b)) and the system remains in a resting state, yielding a stable output (Figure \ref{temperatureres}(c)). As resistance across the thermistor is decreased (temperature increased), controlling the temperature applied to the FBG sensor towards 9.5\,k$\Omega$ (corresponding to a temperature of 26.16\,\textdegree C, red dashed line), the FBG's Bragg wavelength moves closer to the laser emission wavelength, reflecting more light up to a maximum of 8\,mW. In parallel, Figures \ref{temperatureres}(c) and \ref{temperatureres}(d) show that the RTD's electrical response evolves from a resting state to a bursting (spike-firing) regime at first as the thermistor's resistance is decreased to 12\,k$\Omega$ (corresponding to a temperature of 20.84\,\textdegree C for which the FBG's reflected optical power reaches a value of 4.40\,mW). This bursting regime is triggered by the onset of the NDR region skewing very close to the biassing voltage, where small levels of noise are enough to excite temporary excitable responses from the system. Finally the pRTD begins a continuous spiking regime (with a firing rate of 11\,MHz), as shown in Figure \ref{temperatureres}(e), when the thermistor's resistance is decreased to 9.5\,k$\Omega$ (corresponding to a temperature of 26.16\,\textdegree C, for which the reflected optical power is measured to be 7.97\,mW). The results in Figure \ref{temperatureres} therefore demonstrate the ability of the pRTD-FBG neuromorphic system to act as a fibre-optic event-based sensor able to detect and encode remote environmental temperature variations with fast light-triggered spike-firing events. Importantly, the change in spiking frequency of the pRTD gives an insight into the magnitude of temperature change, with increasing temperature yielding slightly faster spike-firing responses at the output of the pRTD (as illustrated in  Figure \ref{rtd characterisation}(b)).

\subsection{Remote Neuromorphic Photonic Event-Based Strain Sensing: Mechanical Perturbations}\label{Strain}

\begin{figure}[h]
    \centering
    \includegraphics[width=\textwidth]{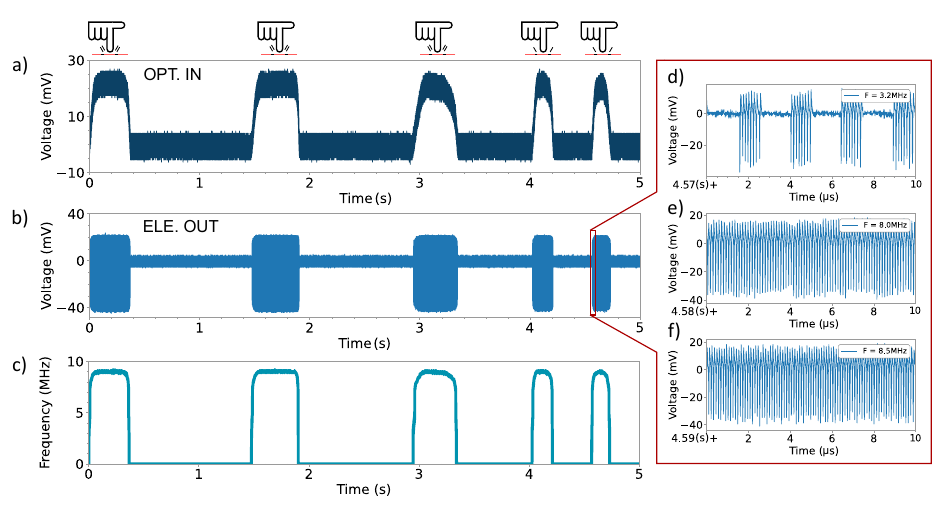}
    \caption{(a) Reflected optical signal from the sensing optical fibre cable with the embedded FBG when perturbed with slow mechanical taps, as measured directly with (a) the standard photodetector included in the setup; (b) with the photonic-electronic pRTD neuron. The latter reveals a spike-firing event-based output in response to detected mechanical strain events. (c) Measurement of the frequency of spiking response from the pRTD over the course of the perturbation applied to the remote sensing optical fibre. (d)-(f) Zoomed insets of a small section of the spiking signal in (c). As reflected power increases, with growing perturbation strength, the system goes from spike bursting with a spike firing frequency of 3.2\,MHz at the onset of spiking, to continuous spiking with 8.5\,MHz frequency.}
    \label{tappingres}
\end{figure}

Now, we probe the system's ability to react to dynamically varying strain-induced events with faster evolving temporal frequency components (up to a few Hz at first). In this configuration, the optical fibre is secured (with tape at one side) to the optical bench and subjected to sweeping mechanical perturbations (finger taps), over the length of the sensor. These stimuli incorporate a sudden mechanical strain onto the FBG embedded within the optical fibre link, temporarily causing a lengthening of its Bragg wavelength. In this way, by setting the tuneable laser wavelength to 1542.250\,nm, with each mechanical tap, the FBG's sensor reflection window is pushed temporarily towards the laser emission wavelength before relaxing back to its unperturbed equilibrium configuration, effectively causing a sudden burst of reflected light. Figure \ref{tappingres} shows experimental results for a 5\,s long demonstration where we manually tap and perturb the system 5 times. Figure \ref{tappingres}(a) shows the perturbed optical signal propagating from the FBG sensor to the pRTD, as measured directly with the amplified photodetector in the setup (see Figure \ref{concept_expsetup}(b)). Sharp increases in detected light indicate the perturbation times.

Figure \ref{tappingres}(b) shows that these are all successfully detected by the neuromorphic photonic event-based detector (pRTD) which responds with fast (ns-rate) electrical spikes (or events), for as long as the strain-induced stimuli are present. A small section of the pRTD dynamical response is highlighted and shown in more detail in Figure \ref{tappingres}(d-f). The latter reveal that similarly to the temperature monitoring case (see Figure \ref{temperatureres}), when detecting a strain variation (finger tap), the RTD dynamical response transitions from an initial quiescent state to a bursting regime at first, and finally to a continuous spike-firing regime. Figure \ref{tappingres}(d) is an example of the initially triggered bursting activity with 3.2\,MHz frequency in the pRTD achieved at the start of the strain event, caused by a low reflected optical power from the FBG, yet significant enough optical injection power bringing the pRTD into the onset of its NDR region. In Figure \ref{tappingres}(e) the reflected optical power is high enough to shift the NDR region to lower voltage that the applied bias to the pRTD falls well into the NDR of its corresponding illuminated I-V characteristic, thereby triggering a continuous spiking pattern (with a measured firing rate of 8\,MHz). Finally, Figure \ref{tappingres}(f) shows that 10\,ms later, the reflected optical power is further increased with rising strain (due to higher alignment of the FBG Bragg wavelength and the laser's wavelength) resulting in the measured spike firing rate to rise to 8.5\,MHz. In addition to the results in Figure \ref{tappingres} obtained with CW light inputs, we also demonstrate (see Section 2 of the supporting information document) the ability of the system to operate with short nano-second-rate pulsed optical inputs (at just 1\% duty cycle); thus, highlighting the potential of the proposed neuromorphic sensing technology to detect rapidly alternating events of interest.

\subsection{Remote Neuromorphic Photonic Event-based Sensing: Audio}

\begin{figure}[h!]
    \centering
    \includegraphics[width=0.9\textwidth]{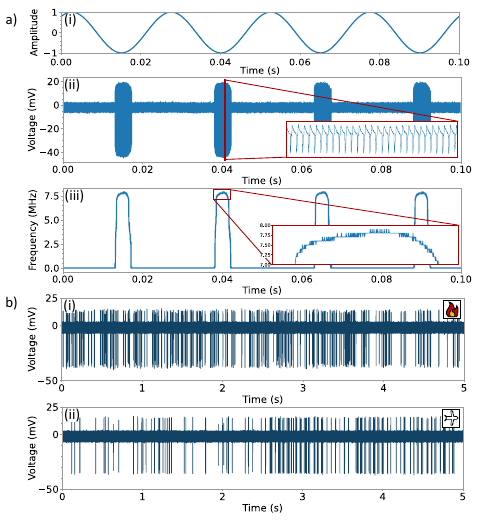}
    \caption{(a)(i) Input sinusoidal waveform  used to create the test audio tone at 40\,Hz, then played through a commercial speaker. (a)(ii)Measured spike-firing response from the pRTD photonic-electronic neuron, firing packets of spikes in accordance with the maximum strain caused by the speaker, with an inset showing the well-defined individual spikes. (a)(iii) The spiking frequency response over time of the pRTD neuron reveals a change in frequency over time with a zoomed inset highlighting the range of frequencies recorded. (b) Measured spiking patterns from the pRTD neuron in response to two test audio files from the ESC-50 dataset. Specifically, files (i) 1-4211-A-12, a crackling fire and (ii) 1-43807-A-47, an aeroplane.}
    \label{audiores}
\end{figure}

Subsequently, we test the ability of the proposed neuromorphic sensing system to also detect faster environmental magnitude changes, focusing initially on audio signals. For this purpose, we attached the optical fibre with the embedded FBG sensor directly to the subwoofer of an off-the-shelf commercial speaker. As audio signals are played through the speaker, the subwoofer vibrates inducing strain within the FBG sensor. At first, we programmed the speaker to emit a 40\,Hz sound tone, as shown in Figure \ref{audiores}(a)(i). Here, troughs induce maximum restrictions of the speaker's diaphragm, causing corresponding periodic peaks of the strain induced within the FBG sensor. These affect dynamically the total reflected light to the pRTD, and the latter yields as a result spike firing responses at instances where maximum strain is induced in the FBG (thus, following the audio tone's frequency) as shown in Figure \ref{audiores}(a)(ii). The inset in Figure \ref{audiores}(a)(ii) highlights the well-defined individual spikes contained within the period of firing activity in the pRTD neuron. The firing rate of the pRTD spiking response is also calculated over the course of the entire measurement in Figure \ref{audiores}(a)(iii) with the zoomed inset highlighting spike rate variations between 7\,MHz and 8\,MHz as the elongation applied by the speaker to the FBG increases. To test the reactivity of the neuromorphic fibre-optic  FBG-pRTD system to audio signals with higher complexity than a simple tone, two distinct sound signals were chosen from the ESC-50: Dataset for Environmental Sound Classification \cite{ESC50} and played through the speaker. The first sound is that of a crackling fire (file 1-4211-A-12) whilst the second one is that of an aeroplane (file 1-43807-A-47). Measured results of the system operation for these two cases are shown respectively in Figure \ref{audiores}(b)(i)\&(ii). These show that over the course of the 5 second temporal duration of the audio files, the amplitude and frequency evolve differently for the two different sound signals, yielding as a result unique (fast) spike firing patterns for each unique sound, upon detection with the pRTD neuron. This leads to the achievement of two different light-triggered event-based signature signals for the two tested sounds, with clearly different spiking patterns. This technique therefore allows not only the capability to detect sounds remotely, but to encode them with fast and event-based (spiking) signals, and offering additional pre-processing capabilities directly at the edge, in the optical domain and at infrared wavelengths. Responses more complex in nature could be potentially achieved by leveraging bistable spiking flip-flop memory\cite{Donati2025b} and future systems with multiple (coupled) RTD neurons \cite{OwenNewns2025b}. This highlights the potentials of the proposed system for its combination with further processing elements to permit identification of specific sounds or other fast varying signals using directly event-based (spiking) produced by the system in response to diverse inputs. 

\subsection{Remote Neuromorphic Photonic Sensing: Fast Turbulence Events}

\begin{figure}[h!]
    \centering
    \includegraphics[width=0.9\textwidth]{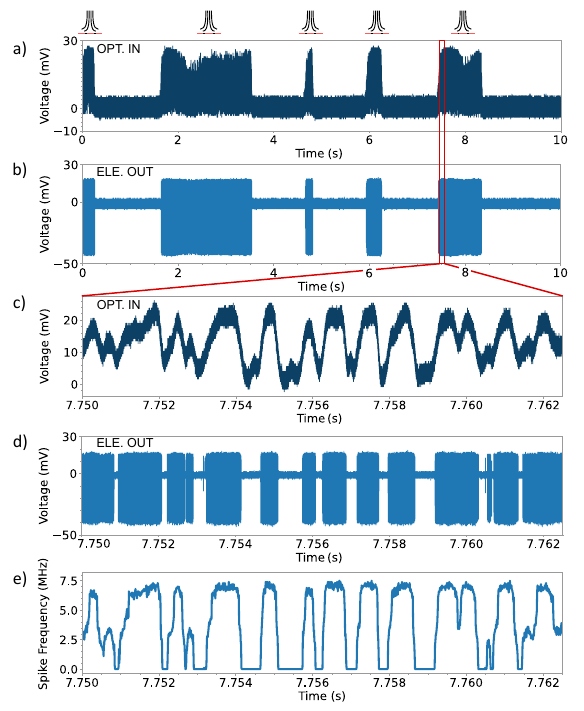}
    \caption{Neuromorphic photonic sensing of remote turbulent events produced with a compressed air duster fired at the optical fibre cable with the embedded FBG sensor. (a) Reflected optical signal from the FBG sensor as directly measured over 10 seconds with the standard photodetector in the setup, showing amplitude variations caused by the turbulent events. (b) Measured spike-firing response from the pRTD neuron in response to the turbulent events. (c) Inset showing a 12.5\,ms section of the signal in (a) highlighting the fast nature of the temporal features of the turbulent signal. (d) 12.5\,ms long sample highlighting the fast spike firing responses delivered by the neuromorphic system. (e) Evolution of the spike firing frequency from the pRTD neuron.}
    \label{turbulenceres}
\end{figure}

We also test the capability of the proposed sensing system to detect even faster events, thanks to the high-speed dynamical spike firing regimes produced by the pRTD neuron at the core of the technology. To do this, we test the system of this work on the detection of fast turbulent events, with frequencies reaching up to a few \,kHz (see results in Figure \ref{turbulenceres}). However, we must note that operation at faster speeds should also be possible with our proposed approach, given the ultrafast spike firing responses achievable in pRTD neurons \cite{Donati2025}. To generate turbulent events, we utilised a compressed air duster, which we brought in close proximity to the optical fibre for 10 seconds, causing turbulent-like dynamical strain events in the FBG sensor embedded in the optical fibre. Figure \ref{turbulenceres}(a) shows the overall optical signal (lasting 10 seconds) reflected from the FBG sensor, as received by the photo-detector in the setup, marking the time locations in which turbulent events are created with the air duster gun. Figure \ref{turbulenceres}(b) shows the full 10 second time-series measured at the output of the pRTD, revealing that this is able to detect all turbulent events by producing continuous spike firing responses for as long as turbulence is present. To highlight the fast dynamic frequencies contained within the input signal a zoomed section of the signal alongside the pRTD response (12.5\,ms-long) is presented in Figures \ref{turbulenceres}(c-e). Specifically, Figure \ref{turbulenceres}(c) shows a zoomed fraction of the optical signal received by the photodetector in Figure \ref{turbulenceres}(a) highlighting its fast temporal characteristics, reaching frequencies of approximately 2\,kHz. Figures \ref{turbulenceres}(d) and \ref{turbulenceres}(e) show respectively the time-series and spike firing rate of the pRTD response, highlighting the dynamical response of the system producing distinct spike firing temporal packets to encode intensity fluctuations within a single turbulence event; thus highlighting the ability of the pRTD to respond and detect fast (turbulent) events. These results further highlight the capability of photonic-electronic spiking pRTDs to serve as fast neuromorphic detectors for (fibre-optic) remote event-based sensing applications in environments subject to fast-changing dynamic events. Here, we note that our proposed approach is able to perform across a very large frequency range, spanning from sub-Hz to \,kHz rates (as shown in this work), but also enabling future operation at much faster rates, thanks to the ultrafast spiking regimes that are possible in RTD neurons \cite{Hejda2022}; hence, opening radically new applications for high speed remote neuromorphic photonic event-based technologies using widely-deployed fibre-optical infrastructure and fully-compatible with infrared optical communication and networking technologies.

\section{Conclusion}\label{Conclusion}

We propose and experimentally demonstrate for the first time to our knowledge, a novel neuromorphic photonic, fibre-optic, event-based and spike-enabled system for remote sensing of events of interest at high speeds. This system combines an optical fibre with embedded FBG sensor operating at key infrared telecom wavelengths with fast photonic-electronic spiking neurons based upon RTD elements. We show that the investigated neuromorphic photonic system can detect (slow-varying) temperature changes, whilst mapping the magnitude of thermal variation to different spike firing rates from the pRTD neuron. Furthermore, we demonstrate the ability of the system to detect and encode faster remote events. This included slow dynamical events induced by strain (at rates in the order of a few Hz) and generated by tapping the fibre directly by touch. Remote event-based detection and encoding of faster audio signals with input frequencies of 10\,s of Hz, was also demonstrated with the pRTD yielding different light-triggered spike firing patterns in response to multiple time-dynamical audio signals. Finally, remote light-enabled, event-based detection of fast turbulent events reaching peak frequencies of approximately 2\,kHz was also demonstrated experimentally. These proof-of-concept demonstrations lay the groundwork for future neuromorphic photonic event-based sensing and in-sensor processing systems for the remote detection of events of interest and fully compatible with infrared fibre-optic telecommunication systems and networks. Notably, whilst in this work demonstrations of fast events up to \,kHz-rate turbulences are reported, the ultrafast operation of the photonic-electronic pRTD spiking neuron at the core of the system are able to deliver ns-rate spiking signals (with theory predicting even faster sub-ns spiking regimes), outlines the potential of this technology for fast remote event-based sensing and monitoring platforms able to operate across a large frequency range. The proposed neuromorphic event-based photonic pRTD-FBG architecture allows multiple routes for future scalability, beyond the initial demonstrations of this work. These may feature future designs incorporating multiple daisy-chained FBG sensors with the same or differing Bragg wavelengths, systems deploying multiple pRTD interrogator elements per FBG sensor or even designs incorporating wavelength, time and space division multiplexing techniques with a combination of FBG and pRTD scaling. These could be deployed in a wide range of complex environments requiring ultrafast detection and monitoring of systems and infrastructure. These may include key societal industry sectors, such as robotics, optical communication networks, civil infrastructure monitoring, perimeter security, among many others.

\section{Methods}\label{Methods}

A continuous wave (CW) laser signal was generated using a tuneable laser (Santec WSL-110, 1527.6-1565.5\,nm) and passed through an optical isolator (to prevent backward reflections) and a polarisation controller before arriving at the input of a Mach-Zehnder intensity modulator (MZM). The MZM was connected to an arbitrary waveform generator (AWG) (312.5\,MSa/s Moku:Pro multi-integrated instrument platform), enabling the system to operate under both CW and pulsed optical input signals (see supporting information document). When operating with CW input, the AWG was disabled. Whilst to operate with optical pulses, RF signals were encoded onto the optical signal using the AWG via the MZM. The output of the MZM was connected to an erbium doped fibre amplifier (EDFA) to amplify the power of the optical signal before its injection in the optical fibre cable used for remote light-enabled sensing operations. Next, the output of the EDFA was connected to port 1 of an optical circulator, steering the signal to port 2, attached to the sensing fibre-optic cable, which contained an embedded FBG sensor (Technica), with a total reflection window of approximately 350\,pm, centred on the peak reflection (Bragg) wavelength of 1541.975\,nm when measured at 295\,K (characterisation can be found within the supporting information document). The wavelength of the optical input and reflected signals was monitored with an optical spectrum analyser (OSA) (Yokogawa AQ6370D 600-1700\,nm), also used to track the FBG resonance wavelength. Varying the temperature or applying strain to the FBG sensor caused its wavelength to detune with respect to the laser injection wavelength, leading to variations in the reflected light reaching a maximum of 89.649\% when the resonance was maximally aligned. In this work, we demonstrate the ability of this system to perform remote detection of a wide range of events, including temperature changes, induced strain and faster environmental changes such as detection of audio and turbulent events. For the measurements concerning temperature detection, the FBG was placed between two metal plates (40x40\,mm) which contained a small access slit, mounted on top of a peltier (ThorLabs TECD6), connected to a 10\,k$\Omega$ thermistor and a temperature controller (ThorLabs TED200C) as shown in Figure \ref{concept_expsetup}(b)(i). For the direct strain detection measures, the optical fibre with the FBG was secured directly onto the optical table and perturbed manually as shown in \ref{concept_expsetup}(b)(ii) for the first round of dynamic measurements (strain-induced), and then attached to the subwoofer of an off-the-shelf commercial speaker, highlighted in Figure \ref{concept_expsetup}(b)(iii) to detect audio signals via the induced vibrations in the fibre by the speaker. Finally, we tested the systems reactivity to turbulent signals by activating a compressed air duster in close proximity to the FBG sensor, shown in Figure \ref{concept_expsetup}(b)(iv). When the CW laser wavelength was aligned with the reflection wavelength of the FBG, the light was reflected to port 2 of the circulator, where the signal was steered to port 3. Port 3 of the circulator was connected to an electrical variable optical attenuator (VOA, ThorLabs V1550A), used for characterisation and further to a 50:50 optical splitter whose first arm was connected to a (ThorLabs PDA8GS) fixed gain amplified Photodetector (PD), whilst the second arm was directly injected to the pRTD via a lens-ended optical fibre. The pRTD was supplied a constant direct current (DC) bias voltage using a Keysight E36312A triple output programmable power supply (PS) through the DC port of a bias-tee. The RF/DC port of the bias-tee was connected via a ground-signal-ground (GSG) probe to the pRTD where the light triggered electrical spiking responses were collected and sent for analysis through the RF port of the tee to a Rohde\&Schwarz RTP084 16\,GHz real time oscilloscope. 

\backmatter

\bmhead{Data Availability}

All data underpinning this publication are openly available from the University of Strathclyde KnowledgeBase at https://doi.org/10.15129/89f64f49-c83c-45f9-8788-23275e4b0f76.

\bmhead{Acknowledgements}

The authors acknowledge support from the European Commission EIC Pathfinder Open project ``SpikePro'' (101129904) and by the UK Research and Innovation (UKRI): Turing AI Acceleration Fellowships Programme (EP/V025198/1), UK Multidisciplinary Centre for Neuromorphic Computing (UKRI982), EPSRC Project ``ProSensing'' (EP/Y030176/1) and the Innovations Knowledge Centre on Neuromorphic Hardware ``NeuroAware''. The authors would also like to acknowledge IQE plc. for providing the semiconductor wafers used to fabricate the systems of this work. For the purpose of open access, the author(s) has applied a Creative Commons Attribution (CC BY) licence to any Author Accepted Manuscript version arising from this submission.

\begin{appendices}

\end{appendices}

\bibliography{sn-bibliography}

\end{document}